\documentclass[review]{elsarticle}
\usepackage{hyperref}
\usepackage{graphicx}
\usepackage{dcolumn}
\usepackage{longtable}
\usepackage{amsmath}
\usepackage{amssymb}

\journal{Nuclear Physics A}
\begin{document}

\begin{frontmatter}

\title
{Quantum phase transitions and collective enhancement of level density  in odd-$A$
and  odd-odd nuclei}

\author[mymainaddress]{S. Karampagia\corref{mycorrespondingauthor}}
\cortext[mycorrespondingauthor]{Corresponding author}
\ead{karampag@nscl.msu.edu}

\author[mysecondaryaddress]{A. Renzaglia}

\author[mymainaddress,mysecondaryaddress]{V. Zelevinsky}

\address[mymainaddress]{National Superconducting Cyclotron Laboratory, Michigan State University, East Lansing, MI 48824-1321, USA}
\address[mysecondaryaddress]{Department of Physics and Astronomy, Michigan State University, East Lansing, MI 48824-1321, USA}

\begin{abstract}
The nuclear shell model assumes an effective mean-field plus interaction Hamiltonian 
in a specific configuration space. We want to understand how various interaction  
matrix elements affect the observables, the collectivity in nuclei and the nuclear 
level density for odd-$A$ and odd-odd nuclei. Using the $sd$ and $pf$ shells, we vary 
specific groups of matrix elements and study the evolution of energy levels, transition 
rates and the level density. In all cases studied, a transition between a ``normal" and 
a collective  phase is induced, accompanied by an enhancement of the level density in 
the collective phase. In distinction to neighboring even-even nuclei, the enhancement 
of the level density is observed already at the transition point. The collective phase 
is reached when the single-particle transfer matrix elements are dominant in the shell 
model Hamiltonian, providing a sign of their fundamental role.

\end{abstract}


\end{frontmatter}

\section{Introduction}

In the framework of the nuclear shell model, an effective Hamiltonian
is used in order to describe the nuclear properties in a certain region of the nuclear
chart. The Hamiltonian can be derived either from
a theory of a deeper level or by a phenomenological fit to experimental data;
in practice one often has to combine these approaches.
The good agreement with the data has rendered the shell model a
powerful tool of nuclear spectroscopy.

The spectroscopic predictions in the framework of the shell model come from
the large-scale diagonalization. Practical necessity to truncate the orbital space
may require the corresponding renormalization of the interaction
and transition operators.
The truncation limits the excitation energy below 
which the shell model predictions can be
reliable (even if we leave aside the continuum decay  thresholds). However, the practically
useful region in many cases already covers the
excitations relevant for  laboratory experiments and for astrophysical reactions.
The shell model also correctly predicts statistical properties of nuclear states.
Therefore it was used as a testing ground for many-body quantum chaos \cite{big}.
In the following, we explore the effects of specific components
of the effective shell-model interactions on the properties of nuclear spectra, and
identify the patterns related to the effects of certain parts of these interactions.
In particular, we study the qualitative changes of nuclear observables
similar to phase transitions which appear as a function of the interaction in the same
shell-model framework. In this way we expect to better understand the relationship
between the input effective Hamiltonian and the nuclear output.

The nuclear level density given by the shell model is sensitive to the specific features of the interaction.
There are successful applications of the shell model to
the prediction of the level density which is a necessary ingredient for the physics of nuclear reactions
\cite{wong,kota,PRC82,PLB11,CPC13,SZ16}.  The traditional Fermi-gas models are based on
the combinatorics of particle-hole excitations
near the Fermi level \cite{fermi1, brody, fermi2}, with the resulting level density
growing exponentially with energy. In order to account for the effects of
pairing  \cite{pairing1, pairing2} or other interactions of
collective nature \cite{collect1, collect2}, various semi-phenomenological or
more elaborate self-consistent mean-field approaches
\cite{Ignatyuk, Bjornholm} have been developed. The shell model Monte Carlo approach, for 
example \cite{Alha}, is close in spirit with the shell model, but may have problems with 
specific interactions and keeping exact quantum numbers. The shell model
Hamiltonian inherently includes pairing and other collective interactions. Along with that,
matrix elements describing incoherent collision-like processes are present as well.
Taking them into account consistently, we come to the level density that, in agreement with
data, is a smooth function of excitation energy. Being still limited by truncated
space, this approach does not require prohibitively large diagonalization. The regular
calculation of the first statistical moments of the Hamiltonian is sufficient for 
reproducing the realistic level density.

In this work we study the evolution of simple nuclear characteristics under the variation of 
the values of certain groups of matrix elements in order to link these matrix elements
to the emergence of collective effects in nuclei. This work can be considered as
an extension of \cite{PRC94} where we limited ourselves to even-even isotopes.
Here we study the behavior of odd-$A$ and odd-odd nuclei in the same mass regions
under the variation of interactions. This provides an additional insight on how the presence
of unpaired fermions affects the changes of nuclear spectral observables and
the level density. As will be seen, the effects of the variation of the matrix elements
in nuclei with unpaired fermions change the nuclear observables in a strong and systematic way.
As a result of the shift of rotational and vibrational excitations to lower energy,
the level density reveals the collective enhancement.

\section{Matrix elements responsible for collectivity}

In the case of the $sd$ shell-model space, there are three single-particle
levels (orbitals), $1s_{1/2}, 0d_{5/2}, 0d_{3/2}$, and 63 matrix elements of the
residual two-body interaction allowed by angular
momentum and isospin conservation. Similarly, for the $pf$ shell, there
are four single-particle levels, $0f_{7/2}, 1p_{3/2}, 0f_{5/2}, 1p_{1/2}$,
and 195 matrix elements of the two-body interaction. The two-body matrix elements
naturally fall into three categories labeled by $\delta=0,1,2$ depending on
how many particles (zero, one or two) change their orbitals as a result of
the interaction process.
We will show that a special role defining the mean-field shape is played by
the ``one unit change", $\delta=1$, matrix elements.

It is known \cite{RM,ZV04} that even a  random (but keeping in force
angular momentum and isospin symmetry) set of matrix elements in a finite orbital
space results in the energy spectrum and properties of stationary states which carry
certain analogies to realistic nuclei. This is essentially a manifestation of the Fermi
statistics and symmetry properties of the orbital space for a given particle number
with  averaging over multiple interaction acts. Artificially changing the reduced
matrix elements $-$ intensifying some interaction processes and weakening others $-$
one can find the interaction landscape  responsible for specific features of individual
 nuclei or their groups.

In a recent study \cite{Horoi} conducted in the shell-model space
$0f_{7/2}, 1p_{3/2}$, the matrix elements allowed in this space were varied randomly
in order to identify those realizations of the  random interaction ensemble
 which give rise to prolate axial deformation.
Among the matrix elements involved, those most important are the $\delta=1$
matrix elements, which are responsible for the mixing of different single-particle
spherical orbitals of the same parity $(|\Delta \ell|=2)$  in the process of quadrupole
deformation. Taking this result into account, the authors in \cite{PRC94} separated the
interaction Hamiltonian into two parts, one containing the
$\delta=1$ matrix elements and another one for the remaining matrix elements.
By varying the relative strength of matrix elements of these groups, a quantum
phase transition was found  in even-even nuclei, both in the $sd$ and $pf$ spaces,
namely a transition  from spherical to deformed shape. The signals of the transition
are the regularities of the lowest yrast energies, including the energy ratio $R_{4/2}$,
the reduced $B$(E2) transition probabilities between these levels, and the amplitudes
of the components of the wave functions.

The deformed phase was realized when the
$\delta=1$ matrix elements  dominated the Hamiltonian, while the spherical phase
arose when the values of these
matrix elements decreased with a simultaneous increase of other matrix elements.
(In a simplified form, similar phase transformations are known in the interacting boson model
\cite{bijker}.) For even-even nuclei, a clear enhancement of the low-energy level density 
was found in the stable deformed phase compared to the spherical phase,  but
not in the vicinity of the transitional point where the shape fluctuations are essential
and various states have complicated wave functions covering both phases. The existence of such fluctuations
measured by the growth of the corresponding correlational entropy of the ground state was found earlier
\cite{VZ03}.
In what follows we explore the effect of the
$\delta=1$ matrix elements in odd-$A$ and odd-odd nuclei in the $sd$ and $pf$ shells.
With the same approach, we will search for signs of a quantum phase transition.

\section{Quantum phase transition}

Nuclear structure models have long provided theoretical tools for
analyzing quantum phase transitions \cite{VZ03,qpt1, qpt2}. Quantum
phase/shape transitions usually occur when the Hamiltonian of the
system is known to have distinct limiting dynamical symmetries
\cite{PT1, PT2, PT3, PT4,PT5,PT6}, revealed in the observables
of the system by varying a control parameter interpolating between the limiting
cases. In the framework of the shell model, pairing and other
collective effects are integrated in the two-body interaction
and the variation of one group of matrix elements with
respect to the others is expected to help  better understand the
role of certain interaction processes in the final properties of  nuclear
observables.

To be sure that our versions of the shell model using the standard values of interaction 
matrix elements are quite realistic, we first demonstrate the quality of the description 
of nuclear data from the results of the full diagonalization
for an odd-odd nucleus $^{26}$Al with rich experimental information, see Table 1. 

\begin{table}

\caption{Experimental energy levels (MeV) and reduced transition probabilities (Weisskopf 
units, W.u.) compared with the shell-model results using the USD interaction \cite{w} for $^{26}$Al.}   \label{table1}
\bigskip
\setlength{\tabcolsep}{8pt}
\begin{center}
\begin{tabular}{ c  c  c  c }
\hline \hline
\multicolumn{2}{c}{Theory} & \multicolumn{2}{c}{Experiment} \\
Energy  & J & Energy  & J \\ \hline
0.000 & 5$^+$ & 0.0 & 5$^+$ \\
0.081 & 0$^+$ & 0.228 & 0$^+$ \\
0.712 & 3$^+$ & 0.417 & 3$^+$ \\
0.819 & 1$^+$ & 1.058 & 1$^+$ \\
1.326 & 2$^+$ & 1.759& 2$^+$ \\
1.737 & 1$^+$ & 1.851& 1$^+$ \\
2.004 & 1$^+$ & 2.069& 4$^+$ \\
2.010 & 2$^+$ & 2.069& 2$^+$ \\
2.121 & 3$^+$ & 2.071& 1$^+$ \\
2.303 & 4$^+$ & 2.365& 3$^+$ \\
2.325 & 3$^+$ & 2.545& 3$^+$ \\ \hline
$B$(E2/M1: $J \rightarrow J'$)  & & $B$(E2/M1: $J \rightarrow J'$) &  \\ \hline
E2: 3$_1^+ \rightarrow 5_1^+$ & 10.61 & E2: 3$_1^+ \rightarrow 5_1^+$ & 8.19 $\pm$ 0.12\\
M1: 1$_1^+ \rightarrow 0_1^+$ & 1.84 & M1: 1$_1^+ \rightarrow 0_1^+$ & 1.5 $\pm$ 0.3\\
E2: 1$_2^+ \rightarrow 3_1^+$ & 10.49 & E2: 1$_2^+ \rightarrow 3_1^+$ & 4.4 $\pm$ 0.8\\
E2: 2$_2^+ \rightarrow 0_1^+$ & 12.93 & E2: 2$_2^+ \rightarrow 0_1^+$ & 12.6  $\pm$ 2.4\\ \hline
\end{tabular}
\end{center}
\end{table}

\begin{figure}
\centering
\includegraphics[height=90mm]{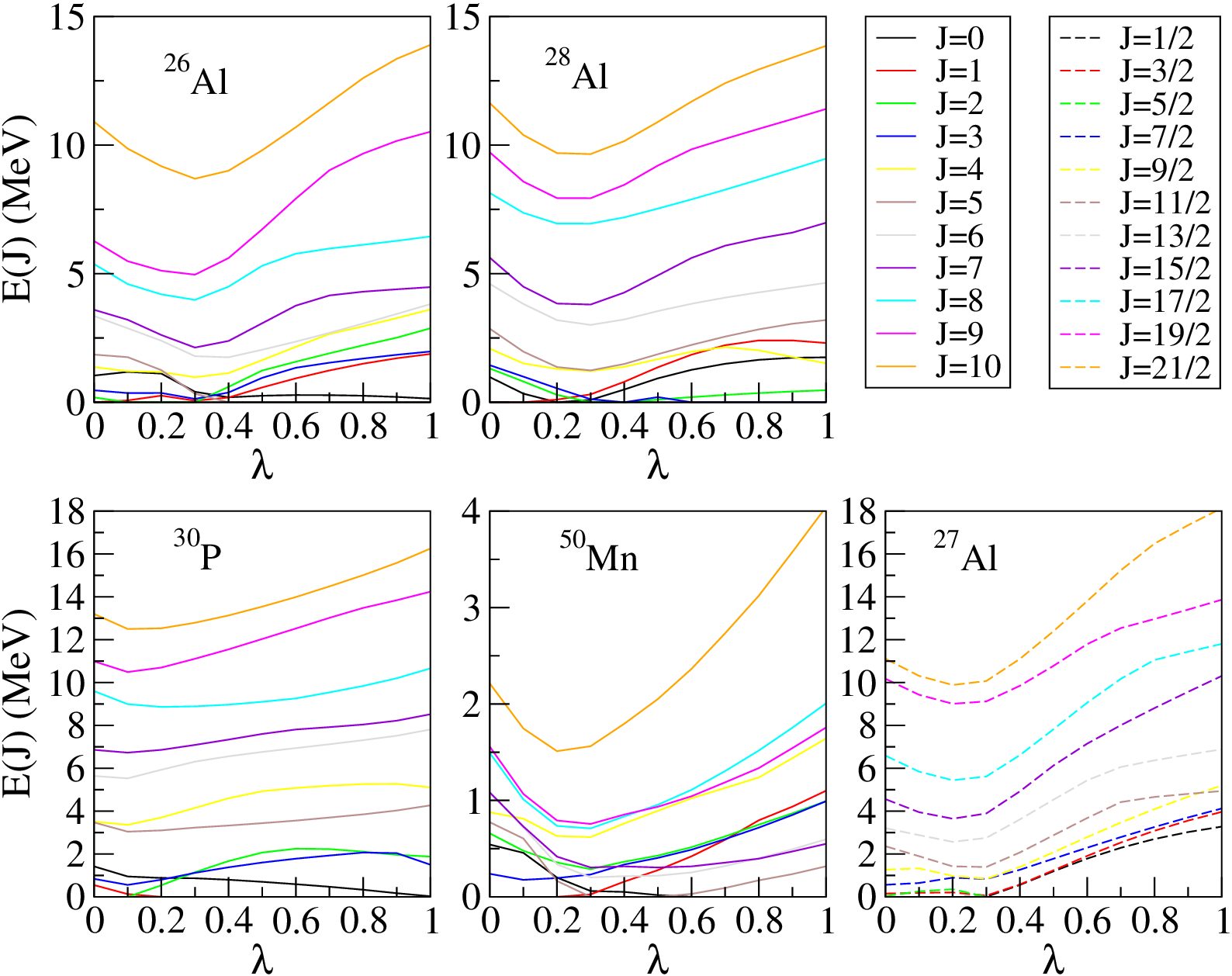}
\caption[]{Yrast energies of $J$=0$-$10 in $^{26,28}$Al, $^{30}$P, $^{50}$Mn. and $J$=1/2$-$21/2 in $^{27}$Al, as a function of 
$\lambda$.}                \label{fig1}
\end{figure}

Simulating the quantum phase transition in the
shell-model framework,  we use now a Hamiltonian of the form,
\begin{equation}
H=h +  (1-\lambda) V_1 +\lambda V_2,  \label{1}
\end{equation}
where $h$ contains the single-particle energies, which will be kept fixed,
and $\lambda$ is the control parameter that varies the values of the
$\delta=1$ matrix elements, $V_1$, and the remaining ($\delta=0$ and 2) matrix
elements, $V_2$. 
Varying $\lambda$  from 0 to 1 in steps of 0.1,
we study the evolution of observables revealed in the chosen $sd$ (odd-odd
and odd-$A$), and $pf$ (odd-odd) nuclei. The results are depicted in
Figs. \ref{fig1}-\ref{fig6}, \ref{fig8} and Tables \ref{table2}-\ref{table7}, 
with Tables \ref{table3}-\ref{table7} found in the Appendix Section.
We have checked that keeping $h$ constant while varying the 
other terms of the Hamiltonian, doesn't affect the qualitative results. 

In even-even nuclei the $\lambda$ dependence
of the energies presents a minimum just for the first few yrast 
low-energy levels. A similar behavior was seen in pairing phase transitions
analyzed through the specially constructed entropy
\cite{VZ03}. 
In contrast to that, here the minimum persists up to high energy values.
For yrast, Fig. \ref{fig1}, as well as for the first ten energy levels
with specific spin values, Fig. \ref{fig2}, there is a clear
minimum of the level energy for all nuclei studied. For all aluminum isotopes
and the $pf$-shell nucleus $^{50}$Mn, the minimum is around $\lambda\cong$ 0.2$-$0.4, while
for $^{30}$P the minimum is at $\lambda\cong$ 0.1. In odd-$A$ and odd-odd nuclei,
depending on the value of $\lambda$, even the ground state spin can change,
showing that the low-energy structure of these nuclei is sensitive to the changes
of the important matrix elements.
This differs from the even-even nuclei, which for the majority of cases
keep the characteristic 0$^+$$-$2$^+$$-$4$^+$ yrast energy sequence
in the process of evolution. The shape phase transition differs from the pairing case
in putting its more efficient imprint up to higher energies. 
% answer spe

To check how the single--particle energies affect this result we repeated the 
calculations after decreasing or increasing the spacings between them. For example, in one case 
we reduced these spacings by a factor 1/2, while in another one we increased them by a factor $1.5$. 
We found that there is always a minimum at a critical 
value of $\lambda$ which persists for all calculated excited states. The displacement  
of single--particle energies affects slightly the position of the minimum of 
energies as a function of $\lambda$. With smaller spacings, the mixing by the $\delta=1$ interaction
is effectively stronger, and the original (supposedly deformed) phase survives longer, the phase transition 
(minimum) appears at larger values of $\lambda$. For instance, in $^{26}$Al 
the minimum of the energies appears closer to $\lambda \cong 0.4$ instead of 
$\lambda \cong 0.3$, while when the single--particle energies are rarefied, the minimum 
appears earlier, for smaller $\lambda$ value ($\lambda \cong 0.2$).  
% end answer spe

\begin{figure}
\centering
\includegraphics[height=90mm]{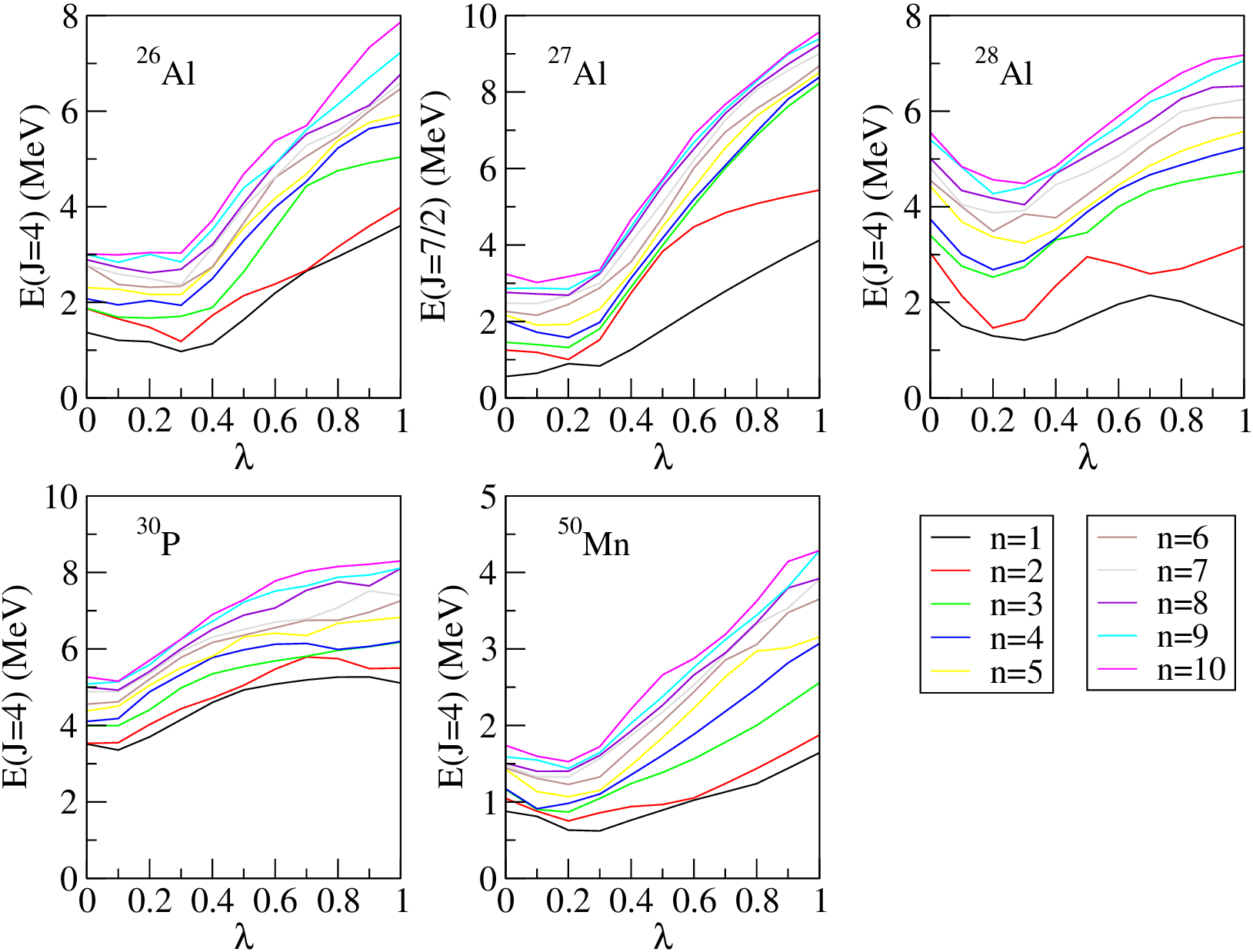}
\caption[]{The first ten energy levels of $J$=4, as a function of 
$\lambda$ for $^{26,28}$Al, $^{30}$P, $^{50}$Mn, and $J$= 7/2 for $^{27}$Al.}\label{fig2}
\end{figure}

Another indicator of the phase transition is the  behavior of the multipole
transition probabilities. In Fig. \ref{fig3} the reduced transition
probabilities $B$(E2;$2^+_1 \rightarrow 0^+_1)$, $B$(E2;$2^+_1 \rightarrow 1^+_1)$,
$B$(E2;$6^+_1 \rightarrow 4^+_1)$ in $^{26,28}$Al, $^{30}$P and
$B$(E2;$(5/2)^+_1 \rightarrow (1/2)^+_1)$, $B$(E2;$(3/2)^+_1 \rightarrow (1/2)^+_1)$,
$B$(E2;$(7/2)^+_1 \rightarrow (3/2)^+_1)$ in $^{27}$Al are presented.
In all cases there is a maximum of the transition rate in the region where the signal of a phase transition 
appears in energies. The probabilities $B$(E2;$2^+_1 \rightarrow 0^+_1)$ for $^{26}$Al
and $B$(E2;$6^+_1 \rightarrow 4^+_1)$ for  $^{30}$P have a maximum at
slightly greater values of $\lambda$.

\begin{figure}[h]
\centering
\includegraphics[height=70mm]{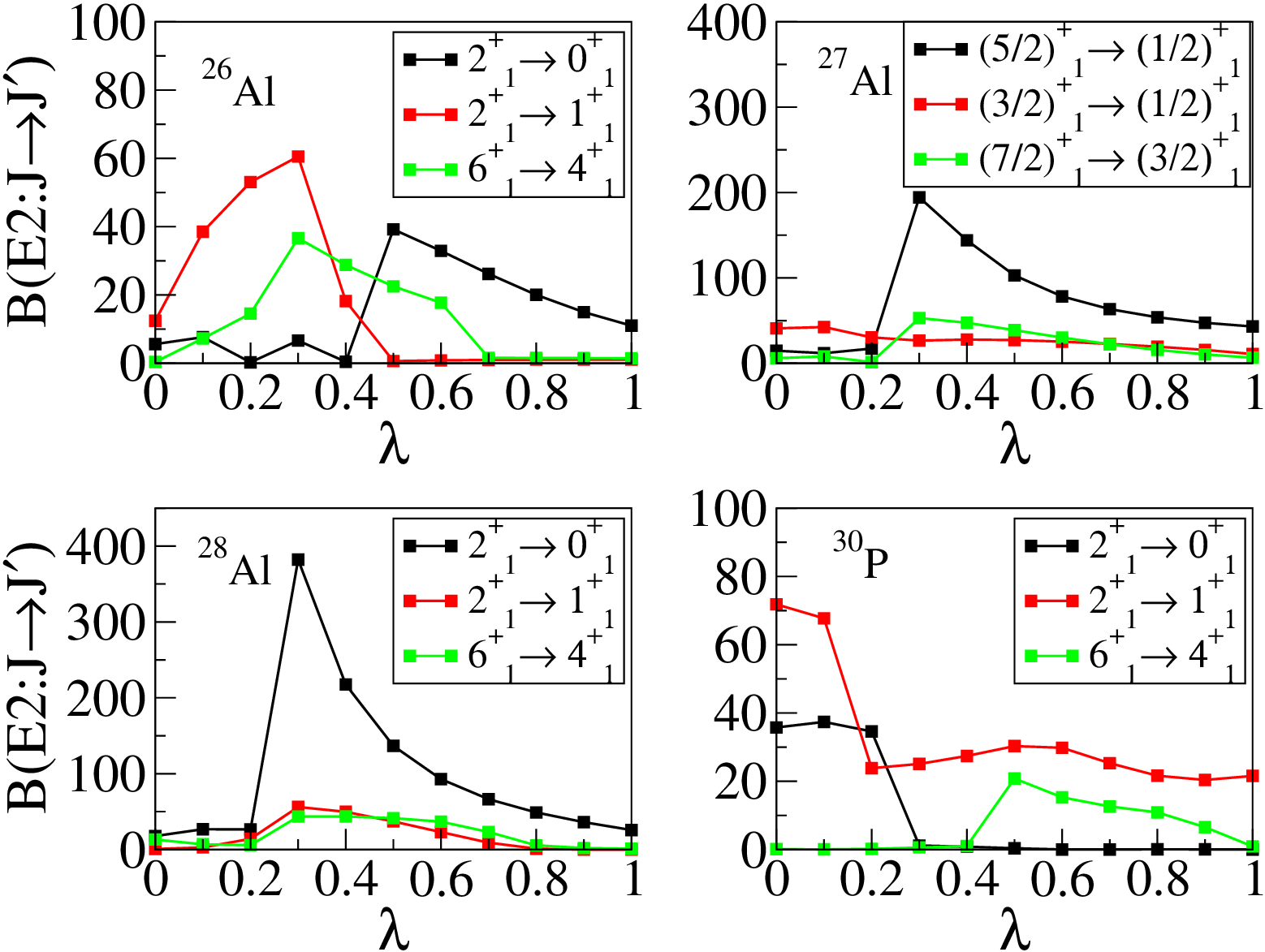}
\caption[]{Reduced quadrupole transition probabilities as a function of 
$\lambda$ for $^{26,27,28}$Al and $^{30}$P.}\label{fig3}
\end{figure}

The proton and neutron spin decomposition of  the wave functions of
different stationary states also presents signs of a quantum phase transition. While this
characteristic extends up to highly excited levels, here we show the decomposition
of the wave function of the $1^+_1$ state that serves as the ground state for
some values of $\lambda$ in all studied nuclei. In Fig. \ref{fig4} we have selected
to show only those components which have an amplitude over 10\%.
Some characteristics are common for all nuclei. First, there is an abrupt change
of the spin decomposition at the transition point. Second, before
the transitional point, there is a strong mixing of the wave function
components, while after the transitional point there are one or two dominant
components, with the rest falling to a minuscule contribution.
For example, for $^{26}$Al and $\lambda \leq 0.2$, the $1^+_1$ state is
mainly made up of protons and neutrons coupled to total angular momentum
according to ($J_n, J_p$) = (5/2, 5/2), (1/2, 1/2), (9/2, 9/2) and (3/2, 5/2) with all
these combinations contributing almost the same. Just
after the transitional point and for  $\lambda > 0.2$, the picture totally
changes. The (5/2, 5/2) combination becomes dominant and stays dominant up
to $\lambda = 1.0$ while other
combinations fall below 5\%.

While the general picture is similar for all three aluminum isotopes, the situation
is slightly different for $^{50}$Mn. Before the transitional point, the ground state of
this nucleus has a strong mixture of
different components, ($J_n, J_p$) = (5/2, 5/2), (7/2, 7/2), (11/2, 11/2),
(7/2, 5/2) and (9/2, 11/2). After the
transitional point the most contributions fall below 10\%, whereas the
two dominant branches persist up to $\lambda = 1.0$. The main
component, again  ($J_n, J_p$) = (5/2, 5/2), stays around 50\%, while
the other one,  ($J_n, J_p$) = (7/2, 7/2), reaches 20\%.

\begin{figure}
\centering
\includegraphics[height=40mm]{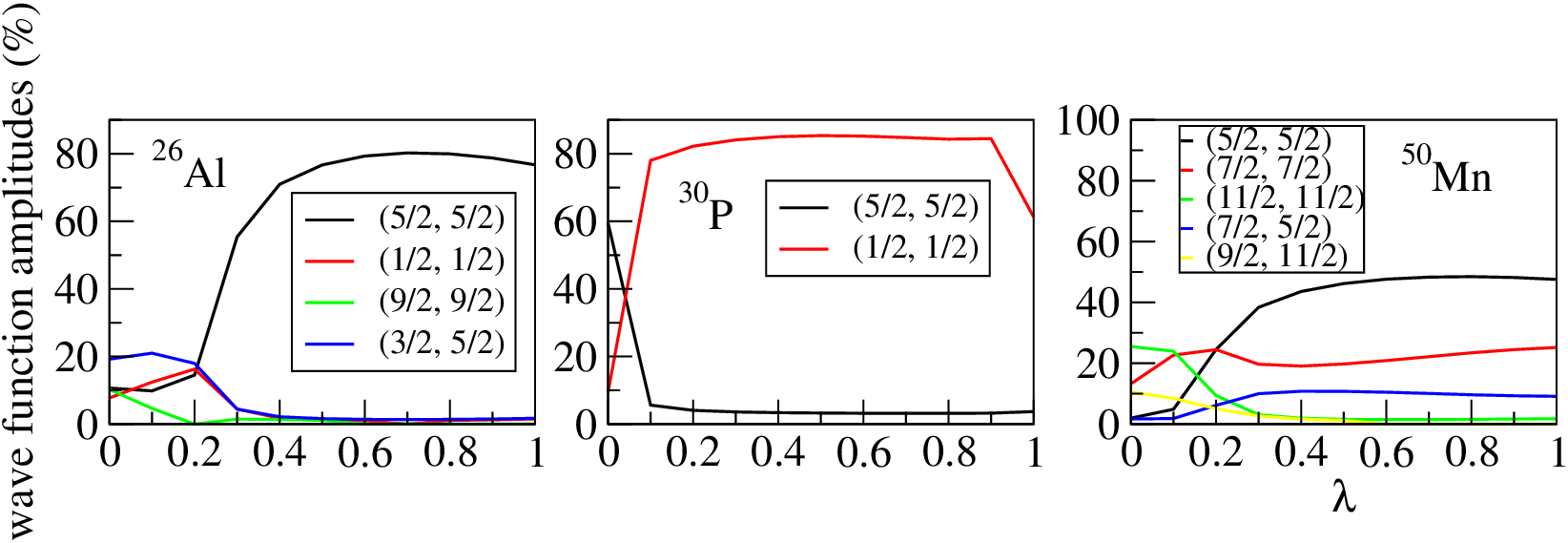}
\caption[]{Amplitudes of the  wave function 1$^+_1$  expanded in terms of proton and neutron
angular momenta as a functions of $\lambda$ for $^{26}$Al, $^{30}$P, and $^{50}$Mn.}\label{fig4}
\end{figure}

The strong mixture of different spin components in the studied wave functions
before the transitional point, compared to the dominance of some spin components
after the transitional point, is related to the occupation of the spherical single-particle
(s.p.) orbitals. Indeed, one can see a small, but observable difference at
the occupation numbers of the s.p. levels in the 1$^+_1$ states for all cases. For
$^{26}$Al and $^{30}$P, just before the transitional point ($\lambda < 0.4$ for
$^{26}$Al and $\lambda < 0.1$ for $^{30}$P), the proton and neutron occupation numbers 
are ($d_{5/2}$, $s_{1/2}$) $\simeq$ (3.5, 1.1) and ($d_{5/2}$, $s_{1/2}$) $\simeq$
(4.6, 1.8) respectively, with the occupation number of $d_{3/2}$ being always less
than 0.5. After the transitional point there is a sudden increase in the occupation
number of $d_{5/2}$, accompanied by a decrease in the occupation of $s_{1/2}$,
($d_{5/2}$, $s_{1/2}$) $\simeq$ (4.6, 0.2) and ($d_{5/2}$, $s_{1/2}$) $\simeq$
(5.8, 1.0) for $^{26}$Al and $^{30}$P, respectively. These changes are relatively small,
but apparently sufficient to induce the mixing characteristics observed in the
wave functions. Similarly, for $^{50}$Mn, the occupation numbers of the s.p. levels
change from ($f_{7/2}$, $p_{3/2}$, $f_{5/2}$) $\simeq$ (4.6, 0.15, 0.15)
before the transition, to ($f_{7/2}$, $p_{3/2}$, $f_{5/2}$) $\simeq$
(4.91, 0.06, 0.03) after the transition, with the $p_{1/2}$ being always less than
0.05. Here from the Nilsson-like occupation scheme we move back to a more normal
spherical shell-model occupancy.

%answer QPT order
In order to find the order of the phase transition we check for discontinuities at the 
first and second derivatives of the ground state energies. We will discuss all the results, 
however we will only show pictures of $^{26}$Al and $^{27}$Al. In the upper panel of 
Fig. \ref{fig5} the ground state energies of $^{26}$Al and  $^{27}$Al are plotted, accompanied 
by the first derivative (middle panel) and second derivative (lower panel) of the ground 
state energy, as a function of $\lambda$. We see that sudden jumps of the first derivative 
of the ground state energy produce emphasized minima in its second derivative. 
Note that here we used steps of 0.01 when moving from $\lambda = 0$ to 1 in order to find 
where the quantum phase transition takes place. The steep minimum of the 
second derivative of $^{27}$Al appears for $\lambda = 0.3$, which is exactly the point where 
we have seen the transitional phenomena. This is analogous for $^{30}$P and 
$^{50}$Mn, where there is a rather steep minimum of the second derivative of the ground 
state energy at  $\lambda = 0.12$ and $\lambda = 0.2$, respectively. We can identify these values of $\lambda$ 
with the phase transition locations.

%Thus, for $^{27}$Al, $^{30}$P 
%and $^{50}$Mn the critical point is at $\lambda = 0.3, 0.12, 0.2$, respectively. 

The situation is slightly different for 
$^{26, 28}$Al, as the second derivative has more than one emphasized minima. These minima 
correspond to the points where spin of the ground state changes. Only the more pronounced minima, 
for example for $^{26}$Al, those at $\lambda = 0.32$ and 0.36, affect the energy behavior, 
inducing two minima at the energies as a function of $\lambda$. These minima are only observable 
at the 0.01 step of $\lambda$. For both nuclei the deepest minimum in the energies comes for the 
largest value of $\lambda$, i.e. for $^{26}$Al at $\lambda = 0.36$ and for $^{28}$Al at 
$\lambda = 0.38$. 

\begin{figure}
\centering
\includegraphics[height=100mm]{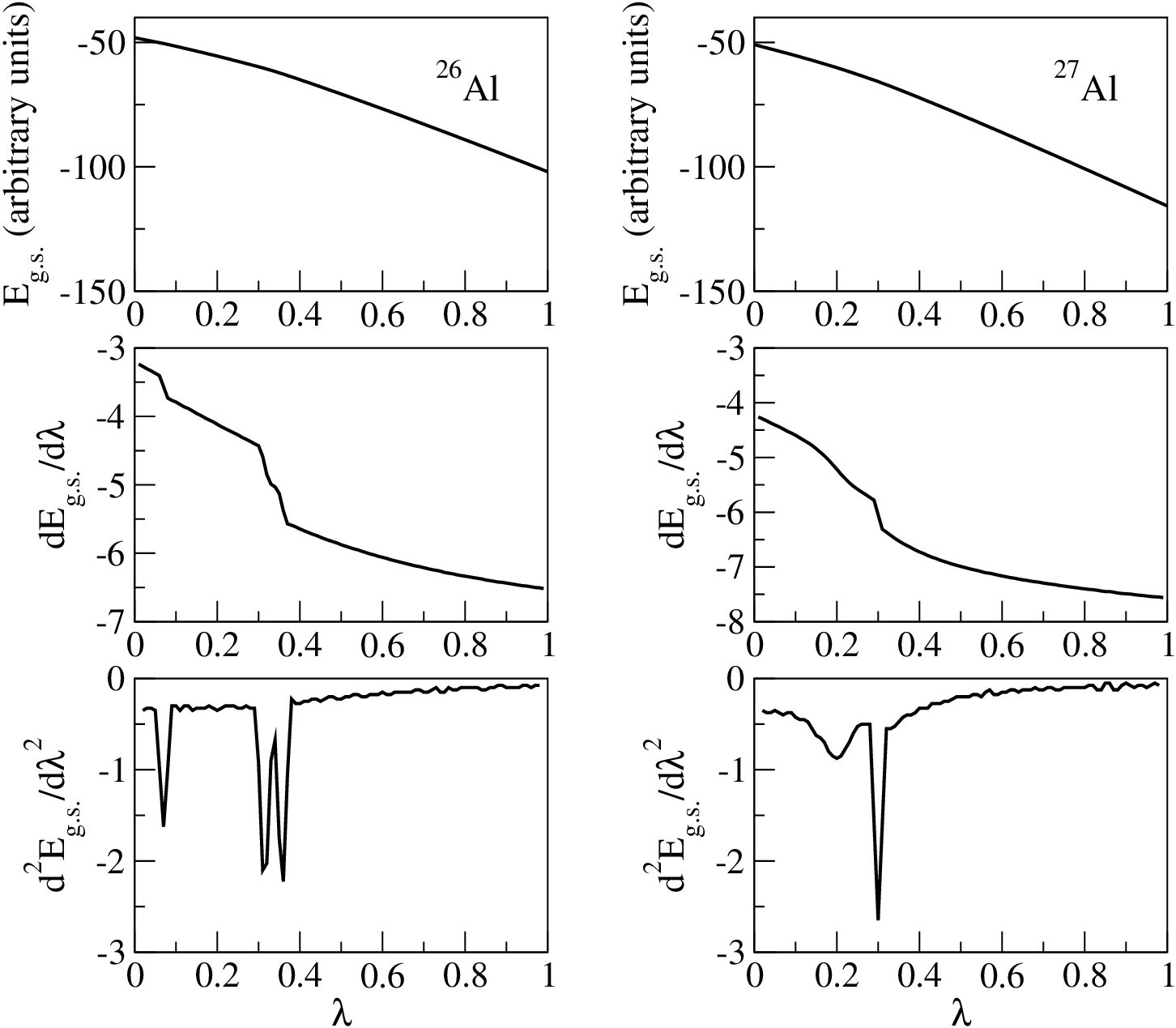}
\caption[]{The ground state energy and its first and second derivatives  for $^{26}$Al and $^{27}$Al as a function 
of $\lambda$.}\label{fig5}
\end{figure}

We look at the wave function, proton and neutron spin decomposition and at the single--particle 
orbital occupancies to understand the structure of the ground state before and after each spike of 
its second derivative. Starting with $^{26}$Al, before the first minimum, which appears at 
$\lambda = $0.07, the wave function consists of 10$\%$ contributions of protons and neutrons 
coupled to angular momenta ($J_n, J_p$) = (5, 5), (1, 1) and (9, 9) and 20$\%$ of (3, 5), (5, 3) components.
The single-particle occupancies are ($d_{5/2}$, $s_{1/2}$) $\simeq$ (3.4, 1.1), the  $d_{3/2}$ 
being less than 0.5. After $\lambda = $0.07 and up to the second minimum, the wave function has 
10$\%$ contribution of pairs of protons and neutrons with spins ($J_n, J_p$) = (5, 5), (3, 5) and 
(5, 3) and 15$\%$ of spins (1, 5), (5, 1). The single-particle occupancies change to ($d_{5/2}$, 
$s_{1/2}$) $\simeq$ (4.0, 0.8), again with a small $d_{3/2}$ component. Thus, the ground state 
is still highly mixed after the first minimum. The situation considerably changes after the 
second minimum. There, the ground state wave function suddenly consists mainly (60$\%$) of 
($J_n, J_p$) = (5, 5), while the $d_{5/2}$ occupation rises to 4.4, leaving the other two 
orbitals with less than 0.5 occupation numbers. After the second minimum the contribution of 
($J_n, J_p$) = (5, 5) protons and neutrons continues to rise reaching 70$\%$, followed by a 
rise to 4.6 of the $d_{5/2}$ occupation number. Therefore, the $\lambda$ value 0.32 is the point 
where the ground state structure changes from mixed to pure configuration.

The case of $^{28}$Al, with two minima at $\lambda =$ 0.17 and 0.28 is similar to $^{26}$Al, the 
ground state wave function changing its character from mixed to pure. Before the first minimum, 
the components of the wavefunction are a mixture of ($J_n, J_p$) = (5, 5), (1, 1) at 20$\%$ and 
($J_n, J_p$) = (1, 3), (5, 3) at 10$\%$, the proton occupation numbers being ($d_{5/2}$, 
$s_{1/2}$) = (3.9, 0.7) and the neutron ($d_{3/2}$, $d_{5/2}$, $s_{1/2}$) = (0.6, 5.2, 1.2). 
Passing the first minimum, only two components of the wave function become important, the ($J_n, 
J_p$) = (5, 5), (1, 1) at 55$\%$ and 33$\%$, respectively, with an accompanying sudden increase 
in the $d_{5/2}$ proton occupancy, ($d_{5/2}$, $s_{1/2}$) = (4.2, 0.6) and the $s_{1/2}$ neutron 
occupancy, ($d_{5/2}$, $s_{1/2}$) = (5.1, 1.5), with the $d_{3/2}$ occupation number falling 
below 0.5. After the second minimum a highly pure ground state is formed, having protons and 
neutrons mainly coupled to ($J_n, J_p$) = (1, 5) at 80$\%$ with the protons mainly occupying the 
$d_{5/2}$ orbital ($d_{5/2}$ = 4.6) and neutrons mainly occupying the $d_{5/2}$ and less the 
$s_{1/2}$ ($d_{5/2}$, $s_{1/2}$ = 5.6, 1.0). 

Trying to see how deformation changes as a function of $\lambda$, we calculate the quadrupole 
moment for $^{26, 28}$Al, $^{30}$P with the results also shown in Fig. \ref{fig6}. It is apparent 
that the details of the interaction change abruplty the quadrupole moment which behaves 
differently  in those three nuclei. However, in all cases, at the point where the ground 
state changes its character from mixed to pure, the quadrupole moment has its maximum value, 
droping to smaller values for $\lambda$ closer to 1. Therefore, the general trend is that, for 
$\lambda$ values closer to one, the deformation is snaller than for $\lambda$ values close to 
zero. 

\begin{figure}
\centering
\includegraphics[height=32mm]{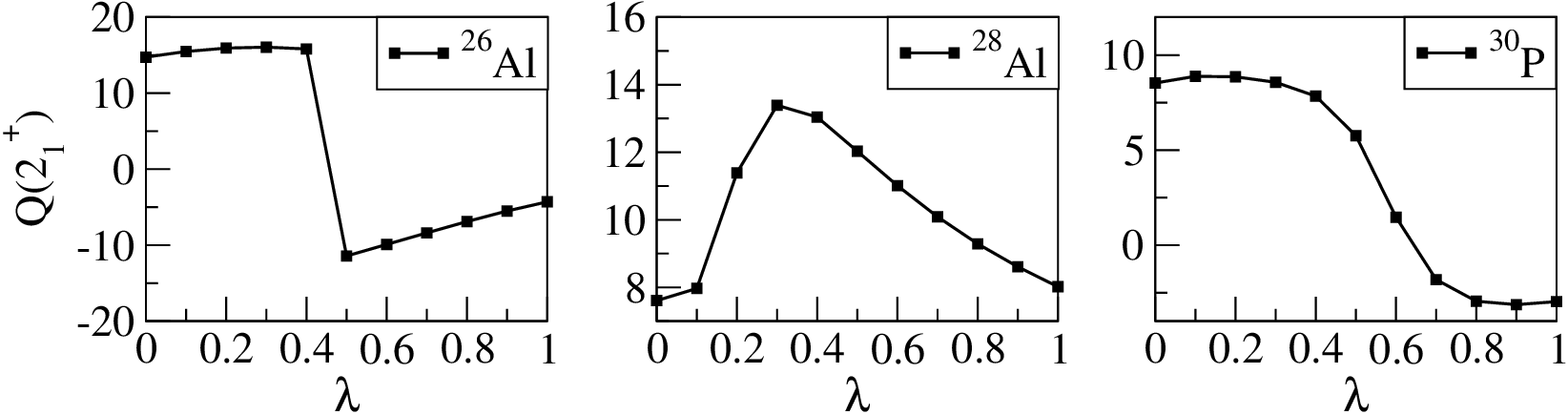}
\caption[]{Quadrupole moments for $^{26}$Al, $^{28}$Al and $^{30}$P as a function of $\lambda$.}\label{fig6}
\end{figure}

Summarizing, the results for the energy levels, multipole transition probabilities, 
the wave function decomposition in proton and neutron spin components, and the quadrupole 
moments reveal the same coherent picture.  At some critical value of $\lambda$, all nuclei undergo 
a transition from a mixed and
collectively deformed phase to a phase close to the 
spherical shape for larger values of $\lambda$. The yrast, and especially
the excited, states of the spectrum show that, for the values of $\lambda$
before the phase transition, the spectrum is overall compressed
to lower energies, while, for larger values of $\lambda$, the spectrum expands to
higher energies. For example, looking at Fig. \ref{fig2}, at $\lambda \leq 0.3$,
the first ten 4$^+$ states of $^{26}$Al have energies below 3 MeV, but for
$\lambda > 0.3$ the 4$^+$ levels shift considerably higher in energy, finally
expanding from 3 to 8 MeV for $\lambda$ = 1.0, when the simple Nilsson-type mixing of
single-particle orbitals is excluded. The study of the derivatives of the ground 
state wave function suggests that this is a second order phase transition. 

There is a principal difference between the nuclear models, mainly algebraic, where 
the quantum phase transitions are studied, and the framework we used to induce a 
quantum phase transition. In the first case, a system is moving between two well 
defined symmetries, while in our case the two groups of matrix elements are not directly
related to any explicit symmetry. The results, though, show clear signs of a qualitative 
change in all studied observables of nuclei, as a function of $\lambda$. There is no
unique critical value of $\lambda$ where this qualitative change takes place, as 
the interaction affects different nuclei  differently. However we clearly see
a coherent behavior of various observables in
different nuclei.    

%end answer QPT order

\section{Level density}

To evaluate the level density up to high excitation energy we use the moments method 
in its current form \cite{CPC13,SZ16} which is based on our knowledge
that the density of states for an individual partition
is close to a Gaussian \cite{wong,kota,brody}. For
a shell-model Hamiltonian that contains a mean-field part and
a residual two-body interaction, the total level density is found by
summing the contributions of all interacting partitions using Gaussians:
\begin{equation}
\rho(E;\alpha)=\sum_{p}D_{\alpha p}G_{\alpha p}(E).                          \label{2}
\end{equation}
In this expression, $\alpha$ combines the quantum numbers of spin, isospin and
parity, while $p$ numbers partitions (various distributions of fermions over
single-particle  orbitals); $D_{\alpha p}$ is the dimension of a given
partition and $G_{\alpha p}$ is a finite-range Gaussian, defined as
\begin{equation}
G_{\alpha p}=G(E-E_{\alpha p}+E_{g.s.};\sigma_{\alpha p}),  \label{3}
\end{equation}
where
\begin{equation}
G(x;\sigma) = C
  \begin{cases}
    e^{-x^2/2\sigma^2},       & |x| \leq \eta \sigma,\\
    0,                        & |x| > \eta \sigma.                            \label{4}
  \end{cases}
\end{equation}
Here, $C$ is the normalizing factor, $\int dx \ G(x;\sigma)=1$, and $\eta$ is
a finite-range cut-off parameter \cite{PRC82}, whose value for this study is 
set to 2.8.  The characteristics of
the finite range Gaussians are determined by the ground state energy
$E_{g.s.}$ and the moments (traces) of the considered Hamiltonian.

For a given partition, the first moment of the Hamiltonian is the centroid, $E_{\alpha p}$, the
mean diagonal matrix element,
\begin{equation}
E_{\alpha p}=\left< H \right>_{\alpha p}=\frac{1}{D_{\alpha p}}{\rm Tr}^{(\alpha p)}H.    \label{5}
\end{equation}
The second moment is the dispersion of the Gaussian, $\sigma_{\alpha p}$,
\begin{equation}
\sigma_{\alpha p}^2=\left< H^2 \right>_{\alpha p}-E_{\alpha p}^2=
\frac{1}{D_{\alpha p}}{\rm Tr}^{(\alpha p)}H^2-E_{\alpha p}^2.
\end{equation}
This is where the mixing of the partitions, due to the interaction processes, 
is accounted for. The calculation of the moments is done directly by the 
Hamiltonian matrix, thus avoiding large matrix diagonalizations.

\begin{figure}[h]
\centering
\includegraphics[height=50mm]{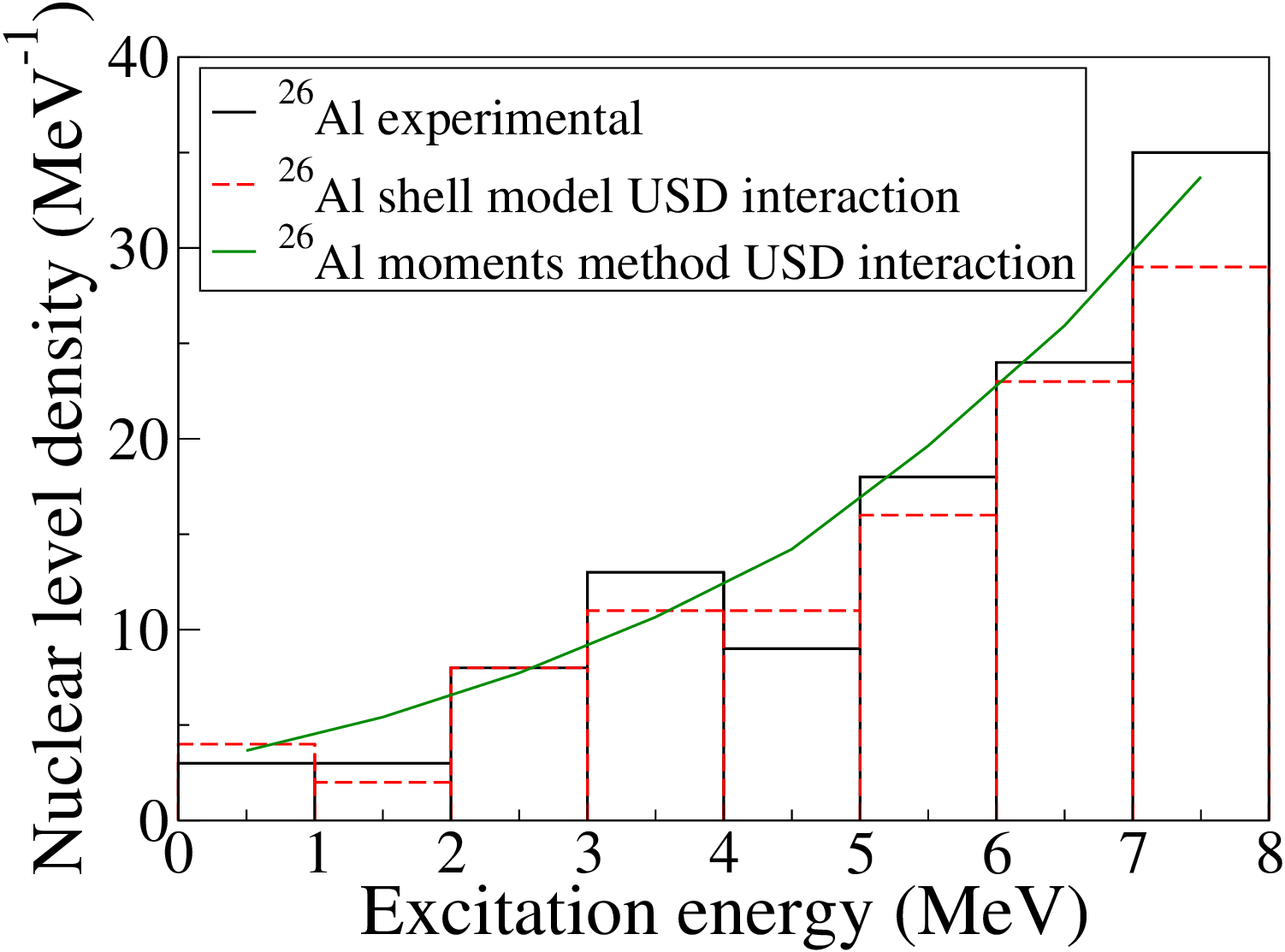}
\caption[]{Comparison of experimental nuclear level density (black stair line) with
the density calculated by full shell-model diagonalization, using the original parameters 
of the interaction (red dashed stair line), and through moments method
(solid green line) for $^{26}$Al, $J$=0$-$7 and positive parity.}\label{fig7}
\end{figure}

The total level density found by the moments method is in good agreement with the results of the
full shell-model calculations. This is illustrated by the example of Fig. \ref{fig7}.
For more details on the moments method,
as well as  comparison with shell model calculations, experimental
results, and Fermi-gas phenomenology, we refer to the previous publications
\cite{PLB11,CPC13,SZ16}.

\begin{figure}
\centering
\includegraphics[height=33mm]{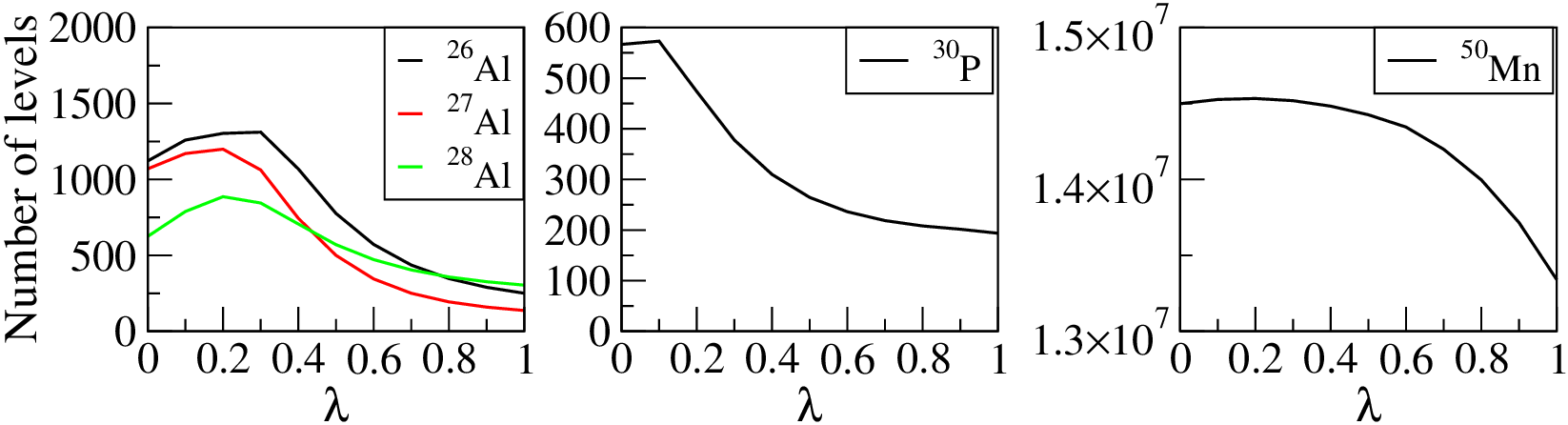}
\caption[]{Number of levels up to 10 MeV for $^{26,28}$Al and $^{30}$P, for $J$=0$-$10
and $J$=1/2$-$21/2 for $^{27}$Al, and up to 60 MeV and $J$=0$-$10 for $^{50}$Mn.}\label{fig8}
\end{figure}

Through the modification of the level density, the  highly excited states  keep memory
of the phase transition that happened
at lower energy and transmitted pronounced effects high along the spectrum. The behavior
of the level density as a function of $\lambda$ is different from the case 
of even-even nuclei \cite{PRC94}. There the level density was falling as a function 
of $\lambda$, though there was a clear enhancement of the level density for 
the cases with the deformed nuclear spectra compared to the vibrational ones. Also, the behavior 
of the level density did not change when considering higher energy states. 
In the current study, the level density increases up to the transitional point and then
decreases strongly till $\lambda$ = 1.0. In Fig. \ref{fig8} and Table \ref{table2},
not only do we observe the enhancement of the level density in the collective phase
of the nuclear system, but we also find the signs of collective enhancement at the
transitional point itself.

The number of levels was calculated using the moments method
for excitation energy up to 10 MeV and spins $J = 0-10$ for $^{26,28}$Al and $^{30}$P,
$J = 1/2-21/2$ for $^{27}$Al, and excitation energy up to 60 MeV and $J = 0-10$ for
$^{50}$Mn. The results, being independent of the angular momentum, do
depend on the selected excitation energy. For example, if one calculates the
cumulative number of levels of the $sd$ nuclei for excitation energy less than 10 MeV,
for some cases (in this study for $^{27}$Al, $^{30}$P), the behavior of the level
density turns out to be different from the one described before. Instead of increasing up
to the transitional point and decreasing after that, the level density 
experiences a continuous decrease as the values of $\lambda$ increase. This decrease,
however, is slow up to the transition and much more abrupt just after the
transition occurs. The situation is similar for $^{50}$Mn where the level density is
continuously decreasing for any excitation energy below 60 MeV,
however the decrease is smooth before the transition and more sharp after that.

\begin{table}

\caption{Cumulative number of levels (NoL) with $J$= 0$-$10 up to energy 10 MeV for different values of $\lambda$
for $^{26,28}$Al, $^{30}$P, with $J$= 1/2$-$21/2 up to energy 10 MeV for $^{27}$Al, and $J$= 0$-$10
up to energy 60 MeV for $^{50}$ Mn.}   \label{table2}
\bigskip
\setlength{\tabcolsep}{8pt}
\begin{center}
\begin{tabular}{ r  r  r  r  r  r  r  r  r  r  r  r }
\hline \hline
$\lambda$& $^{26}$Al & $^{27}$Al & $^{28}$Al & $^{30}$P &$^{50}$Mn \\ \hline

0.0 & 1122 & 1069 & 625 & 567 & 1.449*10$^{7}$ \\
0.1 & 1260 & 1170 & 789 & 573 & 1.452*10$^{7}$ \\
0.2 & 1303 & 1199 & 887 & 474 & 1.453*10$^{7}$ \\
0.3 & 1311 & 1062 & 845 & 378 & 1.451*10$^{7}$ \\
0.4 & 1069 & 744 & 706 & 310 & 1.448*10$^{7}$ \\
0.5 & 776 & 500 & 570 & 265 & 1.442*10$^{7}$ \\
0.6 & 572 & 345 & 472 & 236 & 1.434*10$^{7}$ \\
0.7 & 436 & 251 & 404 & 219 & 1.419*10$^{7}$ \\
0.8 & 348 & 194 & 358 & 208 & 1.399*10$^{7}$ \\
0.9 & 290 & 159 & 326 & 202 & 1.371*10$^{7}$ \\
1.0 & 250 & 137 & 304 & 194 & 1.334*10$^{7}$ \\ \hline
\end{tabular}
\end{center}
\end{table}

In this study, the phase transition is not limited to the ground state and the 
first few excited levels, but it persists up to the very end of the calculated spectrum. The 
persistence of the signs of the phase transition from the ground state up to high 
excitation energy, revealed from the behavior of the excited energy levels and 
the level density, indicates the proliferation of signatures of  the quantum phase transition
beyond the ground state. In distinction to the pairing phase transition that is very clear 
in the ground and pair-vibration states \cite{VZ03} but disappears or becomes a very smooth crossover
in excited states of a small Fermi-system \cite{HZ07}, here many excited states evolve
similarly to the ground state showing essentially the restructuring of the whole mean field. The 
extension of the quantum phase 
transition description to high degrees of excitation  has been under extensive 
research for various many-body models \cite{EQP1, EQP2, EQP3, EQP4, EQP5}. This 
is the first indication of an excited quantum phase transition in the framework of 
the shell model.

\section{Discussion}

In this paper, we studied the evolution of the nuclear observables under the
variation of the values of the matrix elements of the shell-model Hamiltonian,
keeping the exact global symmetries unchanged. Using the two-body residual interaction,
we divided the Hamiltonian into two parts, one containing the ``one unit change",
$\delta=1$, matrix elements, and one containing the rest of two-body matrix elements.
By varying the entrees of the first group in counterphase to the others, we search
for the resulting  behavior of  observables in various nuclei.

The nuclei studied were odd-$A$ ($^{27}$Al) and odd-odd, with either the same
 ($^{26}$Al, $^{30}$P, $^{50}$Mn) or different ($^{28}$Al) number of valence protons and neutrons. 
We concentrated on the signals  of evolution in the energy spectrum and
transition probabilities, the structure of the  stationary wave functions, and the level density in order to search for
the signs of coherent behavior dictated by the variation of the effective many-body Hamiltonian.

Earlier \cite{PRC94} the similar instruments were applied to even-even
nuclei where it was found that a quantum phase transition occurs in the structure of
the first yrast levels, namely the transformation between
rotational and vibrational phases as it was possible to conclude from the evolution
of typical observables. In the current case, a transition between a collective and a
non-collective phase is more pronounced. The $\delta=1$ matrix elements
are indeed carriers of collectivity, acting more strongly on unpaired fermions. As a result,
this transition extends up to the whole spectrum, providing evidence of the
collective enhancement in the level density.
This is seen already at the transitional
point being preformed by the unpaired and freely interacting particles. In this group
of nuclei,  the first example
of an excited-state quantum phase transition
is found in the shell-model framework.

From a slightly more general viewpoint, we open the door into the ``kitchen" of 
the large-scale shell-model diagonalization where usually only the final results
are discussed and compared to the experiment, while the interplay of different 
trends remains hidden in the computations. Meanwhile, looking at the role of individual
players representing various physical components of the interacting system can be
a useful additional source of information about many-body physics. \\
\\

\section*{Acknowledgments}
The work was supported by the NSF grant PHY-1404442. We are thankful 
to A. Volya, B. A. Brown and M. Caprio for numerous discussions.

\newpage

\section*{Appendix A. Tables}

\begin{table}[h]

\caption{Yrast energies of $0^+$$-$$10^+$ (MeV) for $^{26}$Al as a function of $\lambda$.}   \label{table3}
\bigskip
\setlength{\tabcolsep}{8pt}
\begin{center}
\begin{tabular}{ r  r  r  r  r  r  r  r  r  r  r  r }
\hline \hline
$\lambda$& $0^+_1$ & $1^+_1$ & $2^+_1$ & $3^+_1$ &$4^+_1$ & $5^+_1$ & $6^+_1$ & $7^+_1$ & $8^+_1$ & $9^+_1$ & $10^+_1$ \\ \hline
0.0  &  1.038  &  0.000  &  0.190  &  0.466  &  1.368  &  1.854  &  3.355  &  3.595  &  5.373  &  6.268  &  10.907  \\
0.1  &  1.182  &  0.075  &  0.000  &  0.357  &  1.206  &  1.754  &  2.877  &  3.206  &  4.598  &  5.485  &  9.861  \\
0.2  &  1.117  &  0.258  &  0.000  &  0.361  &  1.177  &  1.251  &  2.391  &  2.616  &  4.198  &  5.114  &  9.176  \\
0.3  &  0.408  &  0.062  &  0.000  &  0.129  &  0.973  &  0.331  &  1.793  &  2.126  &  3.979  &  4.962  &  8.691  \\
0.4  &  0.191  &  0.183  &  0.599  &  0.379  &  1.135  &  0.000  &  1.749  &  2.384  &  4.497  &  5.609  &  9.005  \\
0.5  &  0.254  &  0.577  &  1.234  &  0.951  &  1.639  &  0.000  &  2.041  &  3.073  &  5.309  &  6.727  &  9.799  \\
0.6  &  0.280  &  0.930  &  1.576  &  1.344  &  2.159  &  0.000  &  2.358  &  3.756  &  5.778  &  7.920  &  10.695  \\
0.7  &  0.278  &  1.238  &  1.905  &  1.538  &  2.656  &  0.000  &  2.700  &  4.155  &  5.977  &  9.022  &  11.649  \\
0.8  &  0.253  &  1.499  &  2.216  &  1.701  &  2.955  &  0.000  &  3.062  &  4.301  &  6.124  &  9.669  &  12.602  \\
0.9  &  0.208  &  1.711  &  2.510  &  1.846  &  3.274  &  0.000  &  3.436  &  4.393  &  6.278  &  10.160  &  13.352  \\
1.0  &  0.144  &  1.876  &  2.786  &  1.973  &  3.607  &  0.000  &  3.817  &  4.477  &  6.448  &  10.520  &  13.896  \\ \hline \hline
\end{tabular}
\end{center}
\end{table}

\begin{table}

\caption{Yrast energies of $1/2^+$$-$$21/2^+$ (MeV) for $^{27}$Al as a function of $\lambda$.}   \label{table4}
\bigskip
\setlength{\tabcolsep}{8pt}
\begin{center}
\begin{tabular}{ r  r  r  r  r  r  r  r  r  r  r  r }
\hline \hline
$\lambda$& ${1/2}^+_1$ & ${3/2}^+_1$ & ${5/2}^+_1$ & ${7/2}^+_1$ &${9/2}^+_1$ & ${11/2}^+_1$ & ${13/2}^+_1$ & ${15/2}^+_1$ & ${17/2}^+_1$ & ${19/2}^+_1$ & ${21/2}^+_1$ \\ \hline
0.0  &  0.000  &  0.152  &  0.038  &  0.562  &  1.269  &  2.365  &  3.209  &  4.559  &  6.593  &  10.177  &  11.112  \\
0.1  &  0.000  &  0.187  &  0.258  &  0.647  &  1.330  &  1.899  &  2.878  &  3.952  &  5.845  &  9.427  &  10.312  \\
0.2  &  0.000  &  0.207  &  0.355  &  0.894  &  0.979  &  1.423  &  2.568  &  3.647  &  5.441  &  9.007  &  9.886  \\
0.3  &  0.003  &  0.059  &  0.000  &  0.836  &  0.842  &  1.396  &  2.752  &  3.889  &  5.616  &  9.118  &  10.069  \\
0.4  &  0.560  &  0.578  &  0.000  &  1.265  &  1.392  &  2.073  &  3.612  &  4.926  &  6.614  &  9.850  &  11.106  \\
0.5  &  1.195  &  1.241  &  0.000  &  1.780  &  2.081  &  2.874  &  4.539  &  6.129  &  7.825  &  10.783  &  12.399  \\
0.6  &  1.787  &  1.911  &  0.000  &  2.295  &  2.787  &  3.678  &  5.436  &  7.149  &  9.060  &  11.791  &  13.801  \\
0.7  &  2.291  &  2.533  &  0.000  &  2.790  &  3.464  &  4.422  &  6.060  &  8.006  &  10.173  &  12.538  &  15.235  \\
0.8  &  2.701  &  3.084  &  0.000  &  3.260  &  4.093  &  4.662  &  6.371  &  8.805  &  11.053  &  12.966  &  16.480  \\
0.9  &  3.025  &  3.560  &  0.000  &  3.705  &  4.669  &  4.804  &  6.632  &  9.568  &  11.439  &  13.403  &  17.342  \\
1.0  &  3.275  &  3.965  &  0.000  &  4.122  &  5.194  &  4.937  &  6.886  &  10.305  &  11.800  &  13.863  &  18.137  \\ \hline \hline
\end{tabular}
\end{center}
\end{table}

\begin{table}

\caption{Yrast energies of $0^+$$-$$10^+$ (MeV) for $^{28}$Al as a function of $\lambda$.}   \label{table5}
\bigskip
\setlength{\tabcolsep}{8pt}
\begin{center}
\begin{tabular}{ r  r  r  r  r  r  r  r  r  r  r  r }
\hline \hline
$\lambda$& $0^+_1$ & $1^+_1$ & $2^+_1$ & $3^+_1$ &$4^+_1$ & $5^+_1$ & $6^+_1$ & $7^+_1$ & $8^+_1$ & $9^+_1$ & $10^+_1$ \\ \hline
0.0 & 0.982 & 0.000 & 1.316 & 1.442 & 2.086 & 2.860 & 4.602 & 5.623 & 8.137 & 9.720 & 11.636 \\
0.1 & 0.334 & 0.000 & 0.809 & 0.996 & 1.511 & 1.969 & 3.824 & 4.493 & 7.365 & 8.580 & 10.394 \\
0.2 & 0.000 & 0.102 & 0.290 & 0.542 & 1.297 & 1.370 & 3.194 & 3.839 & 6.951 & 7.940 & 9.690 \\
0.3 & 0.094 & 0.307 & 0.000 & 0.119 & 1.211 & 1.237 & 3.007 & 3.801 & 6.946 & 7.936 & 9.651 \\
0.4 & 0.495 & 0.795 & 0.000 & 0.001 & 1.377 & 1.489 & 3.219 & 4.264 & 7.190 & 8.454 & 10.154 \\
0.5 & 0.929 & 1.361 & 0.105 & 0.201 & 1.678 & 1.868 & 3.546 & 4.936 & 7.532 & 9.205 & 10.901 \\
0.6 & 1.262 & 1.849 & 0.201 & 0.000 & 1.964 & 2.229 & 3.831 & 5.612 & 7.889 & 9.834 & 11.693 \\
0.7 & 1.498 & 2.218 & 0.285 & 0.000 & 2.148 & 2.555 & 4.068 & 6.090 & 8.265 & 10.243 & 12.408 \\
0.8 & 1.650 & 2.404 & 0.358 & 0.000 & 2.018 & 2.836 & 4.273 & 6.373 & 8.659 & 10.630 & 12.941 \\
0.9 & 1.729 & 2.403 & 0.418 & 0.000 & 1.763 & 3.055 & 4.459 & 6.595 & 9.063 & 11.012 & 13.400 \\
1.0 & 1.747 & 0.880 & 0.470 & 0.000 & 1.514 & 3.198 & 4.643 & 6.980 & 9.473 & 11.403 & 13.856 \\
 \hline \hline
\end{tabular}
\end{center}
\end{table}

\begin{table}

\caption{Yrast energies of $0^+$$-$$10^+$ (MeV) for $^{30}$P as a function of $\lambda$.}   \label{table6}

\bigskip
\setlength{\tabcolsep}{8pt}
\begin{center}
\begin{tabular}{ r  r  r  r  r  r  r  r  r  r  r  r }
\hline \hline
$\lambda$& $0^+_1$ & $1^+_1$ & $2^+_1$ & $3^+_1$ &$4^+_1$ & $5^+_1$ & $6^+_1$ & $7^+_1$ & $8^+_1$ & $9^+_1$ & $10^+_1$ \\ \hline
0.0  &  1.418  &  0.546  &  0.000  &  0.839  &  3.531  &  3.473  &  5.643  &  6.861  &  9.604  &  10.986  &  13.200  \\
0.1  &  0.947  &  0.127  &  0.000  &  0.559  &  3.357  &  3.036  &  5.528  &  6.730  &  8.986  &  10.491  &  12.497  \\
0.2  &  0.879  &  0.000  &  0.544  &  0.804  &  3.700  &  3.097  &  5.929  &  6.857  &  8.863  &  10.698  &  12.526  \\
0.3  &  0.865  &  0.000  &  1.152  &  1.119  &  4.151  &  3.225  &  6.306  &  7.089  &  8.892  &  11.105  &  12.790  \\
0.4  &  0.798  &  0.000  &  1.673  &  1.380  &  4.600  &  3.331  &  6.561  &  7.340  &  8.968  &  11.554  &  13.134  \\
0.5  &  0.701  &  0.000  &  2.065  &  1.596  &  4.927  &  3.439  &  6.755  &  7.597  &  9.101  &  12.031  &  13.539  \\
0.6  &  0.590  &  0.000  &  2.250  &  1.776  &  5.079  &  3.560  &  6.935  &  7.811  &  9.292  &  12.524  &  13.993  \\
0.7  &  0.469  &  0.000  &  2.229  &  1.932  &  5.189  &  3.699  &  7.118  &  7.920  &  9.541  &  13.023  &  14.485  \\
0.8  &  0.333  &  0.000  &  2.110  &  2.074  &  5.264  &  3.854  &  7.311  &  8.042  &  9.843  &  13.478  &  15.013  \\
0.9  &  0.172  &  0.000  &  1.962  &  2.036  &  5.265  &  4.027  &  7.519  &  8.221  &  10.198  &  13.838  &  15.581  \\
1.0  &  0.030  &  0.000  &  1.878  &  1.502  &  5.107  &  4.274  &  7.807  &  8.520  &  10.662  &  14.238  &  16.252  \\ \hline \hline
\end{tabular}
\end{center}
\end{table}

\begin{table}

\caption{Yrast energies of $0^+$$-$$10^+$ (MeV) for $^{50}$Mn as a function of $\lambda$.}   \label{table7}
\bigskip
\setlength{\tabcolsep}{8pt}
\begin{center}
\begin{tabular}{ r  r  r  r  r  r  r  r  r  r  r  r }
\hline \hline
$\lambda$& $0^+_1$ & $1^+_1$ & $2^+_1$ & $3^+_1$ &$4^+_1$ & $5^+_1$ & $6^+_1$ & $7^+_1$ & $8^+_1$ & $9^+_1$ & $10^+_1$ \\ \hline
0.0  &  0.543  &  0.000  &  0.658  &  0.239  &  0.877  &  0.774  &  1.087  &  1.083  &  1.493  &  1.558  &  2.213  \\
0.1  &  0.456  &  0.000  &  0.478  &  0.176  &  0.810  &  0.605  &  0.720  &  0.729  &  1.011  &  1.067  &  1.746  \\
0.2  &  0.200  &  0.000  &  0.357  &  0.195  &  0.631  &  0.165  &  0.337  &  0.417  &  0.735  &  0.792  &  1.511  \\
0.3  &  0.061  &  0.026  &  0.287  &  0.232  &  0.620  &  0.000  &  0.202  &  0.305  &  0.710  &  0.756  &  1.562  \\
0.4  &  0.051  &  0.158  &  0.364  &  0.335  &  0.761  &  0.000  &  0.214  &  0.316  &  0.834  &  0.853  &  1.795  \\
0.5  &  0.018  &  0.277  &  0.428  &  0.404  &  0.892  &  0.000  &  0.219  &  0.305  &  0.955  &  0.934  &  2.051  \\
0.6  &  0.000  &  0.421  &  0.515  &  0.489  &  1.023  &  0.031  &  0.253  &  0.316  &  1.110  &  1.043  &  2.364  \\
0.7  &  0.000  &  0.592  &  0.627  &  0.599  &  1.129  &  0.092  &  0.320  &  0.355  &  1.305  &  1.188  &  2.733  \\
0.8  &  0.000  &  0.794  &  0.749  &  0.718  &  1.238  &  0.171  &  0.401  &  0.396  &  1.517  &  1.335  &  3.122  \\
0.9  &  0.000  &  0.935  &  0.866  &  0.849  &  1.438  &  0.236  &  0.493  &  0.471  &  1.754  &  1.543  &  3.576  \\
1.0  &  0.000  &  1.101  &  0.993  &  0.992  &  1.642  &  0.317  &  0.597  &  0.550  &  2.006  &  1.756  &  4.042  \\ \hline \hline
\end{tabular}
\end{center}
\end{table}

\end{document}